\documentstyle[twocolumn,eqsecnum,aps]{revtex}

\newcommand{\nc}{\newcommand}
\nc{\be}{\begin{equation}}
\nc{\ee}{\end{equation}}
\nc{\bea}{\begin{eqnarray}}
\nc{\eea}{\end{eqnarray}}
\nc{\disp}{\displaystyle}
\nc{\ade}{\mbox{$A$-$D$-$E$}}
\nc{\calN}{{\cal N}}
\nc{\calC}{{\cal C}}
\nc{\calM}{{\cal M}}
\nc{\calS}{{\cal S}}
\nc{\phit}{\hat{\varphi}}
\nc{\chit}{\hat{\chi}}
\nc{\hcalN}{\hat{\calN}}
\nc{\hcalS}{\hat{\calS}}
\nc{\hS}{\hat{S}}
\nc{\sigmad}{\sigma^\dagger}
\nc{\psid}{\psi^\dagger}
\def\smat#1{\mbox{\scriptsize{\mbox{$\pmatrix{#1}$}}}}

\font\tenmsb=msbm10
\font\sevenmsb=msbm7
\font\fivemsb=msbm5
\newfam\msbfam
\textfont\msbfam=\tenmsb
\scriptfont\msbfam=\sevenmsb
\scriptscriptfont\msbfam=\fivemsb
\def\Bbb#1{{\fam\msbfam\relax#1}}
\def\half {\mbox{$\textstyle {1 \over 2}$}}

\def\punit#1{\hspace{#1\unitlength}}

\def\vert#1#2#3{\rule[-4\unitlength]{0in}{8\unitlength}
\begin{picture}(0,0)(-#1,-#2)
\put(0,0){\line(0,1){4}}
\put(0,-1){\makebox(0,0)[t]{\smaller \mbox{$#3$}}}
\end{picture}}

\def\vertl#1#2#3#4{
\begin{picture}(0,0)(-#1,-#2)
\put(0,0){\line(0,1){#4}}
\put(0,-1){\makebox(0,0)[t]{\smaller \mbox{$#3$}}}
\end{picture}}

\def\monoid#1#2#3#4{\rule[-4\unitlength]{0in}{8\unitlength}
\begin{picture}(0,0)(-#1,-#2)
\put(2,0){\oval(4,3.5)[t]}
\put(2,4){\oval(4,3.5)[b]}
\put(0,-1){\makebox(0,0)[t]{\smaller \mbox{$#3$}}}
\put(4,-1){\makebox(0,0)[t]{\smaller \mbox{$#4$}}}
\end{picture}}

\def\downmonoid#1#2#3#4{\rule[-0\unitlength]{0in}{1\unitlength}
\begin{picture}(0,0)(-#1,-#2)
\put(2,0){\oval(4,3.5)[t]}
\put(0,-1){\makebox(0,0)[t]{\smaller \mbox{$#3$}}}
\put(4,-1){\makebox(0,0)[t]{\smaller \mbox{$#4$}}}
\end{picture}}

\def\down2monoid#1#2#3#4{\rule[-4\unitlength]{0in}{8\unitlength}
\begin{picture}(0,0)(-#1,-#2)
\put(2,0){\oval(12,10)[t]}
\put(-4,-1){\makebox(0,0)[t]{\smaller \mbox{$#3$}}}
\put(8,-1){\makebox(0,0)[t]{\smaller \mbox{$#4$}}}
\end{picture}}

\def\smaller{\tiny}
\def\bra#1{\langle #1|}
\def\ket#1{|#1\rangle}

\begin{document}
\draft
\title{Critical $Q=1$ Potts Model\\
and Temperley-Lieb Stochastic Processes
}
\author{Paul A. Pearce, 
Vladimir Rittenberg 
}
\address{Department of Mathematics and Statistics\\
University of Melbourne, 
Parkville, Victoria 3010, Australia}
\author{Jan de Gier 
}
\address{Department of Mathematics, School of Mathematical Sciences\\
Australian National University, Canberra ACT 0200, Australia}

\date{\today}
\maketitle
\widetext
\begin{abstract}
We consider the groundstate wave function and spectra of the $L$-site XXZ $U_q[s\ell(2)]$
invariant quantum spin chain with  $q=\exp(\pi i/3)$. This chain is related to
the critical $Q=1$ Potts model and exhibits $c=0$ conformal invariance. 
We show that the problem is related to
Hamiltonians describing one-dimensional stochastic processes defined
on a Temperley-Lieb algebra. The bra groundstate wave function is
trivial and the ket groundstate wave function gives the probabilty
distribution of the stationary state. The stochastic processes can be
understood as interface RSOS growth models with nonlocal rates. 
Allowing defects which can hop on the
interface one obtains stochastic models having the same
stationary state and spectra (but not degeracies) as the XXZ chain. 

\end{abstract}

\narrowtext

\section{INTRODUCTION}
\label{intro}

It is known that the $Q=1$ Potts model~\cite{Mart91} is 
related to other models in statistical mechanics such as percolation~\cite{perc}
and the $O(1)$ loop model~\cite{BaxtKW76}. It was only
recently realized, however, that at criticality it is related to other
topics in physics and mathematics. Consider the
ferromagnetic quantum $U_q[s\ell(2)]$ invariant XXZ
Hamiltonian~\cite{PasqS90}
\bea
&&\hspace{.5in}H_{\text{XXZ}}=\sum_{j=1}^{L-1}(1-e_j)\label{Ham}\\
e_j&=&{1\over 4}-{1\over2}\left[\Big(\sigma^x_j\sigma^x_{j+1}+\sigma^y_j\sigma^y_{j+1}
+\Delta\sigma^z_j\sigma^z_{j+1}\Big)
\right.\nonumber\\&&\left.\mbox{}
+h\Big(\sigma^z_j-\sigma^z_{j+1}\Big)\right]
\eea
where $\sigma^x,\sigma^y,\sigma^z$ are Pauli matrices, $q=e^{\pi i/3}$ and
\be
2\Delta=q+q^{-1}=\sqrt{Q}=1,\quad h=\half(q-q^{-1})=i\mbox{${\sqrt{3}\over 2}$}
\ee
The value $\Delta=1/2$ relates to the critical $Q=1$ Potts model. 
The $e_j$ are the generators of a Temperley-Lieb (TL) algebra
\be
e_j^2=e_j,\quad e_je_{j\pm 1}e_j=e_j,
\quad [e_i,e_j]=0,\  |i-j|>1.
\ee
The  groundstate energy is zero, that is $H_{\text{XXZ}}\ket0=0\ket0$, for
any number of sites as can be seen by taking the quotient
$e_j=1$ of the TL algebra. 

As pointed out by Read and Saleur~\cite{ReadS01}
the spectrum of the Hamiltonian (1.1) is also related, in\, the\, continuum
\newpage
\mbox{}
\vspace{1.7in}
\mbox{}

\noindent
limit, to the spectra of  nonlinear sigma models defined on coset
supermanifolds. Such models  are relevant in the integer quantum Hall
effect~\cite{GruzLR99} and problems with quenched 
disorder~\cite{GuruLL00}. The spectra of all of these
models are given by a generalization of the $c=0$ Virasoro algebras related
to logarithmic conformal field theories (LCFTs)~\cite{GuraL99} (for reviews
of the subject see \cite{Floh00}). What is common to all
these theories is the Jordan cell structure of the representations which
occur when studying their spectra. This is due to the fact that, at
$q$ a root of unity, indecomposable representations of $U_q[s\ell(2)]$ 
appear and the same is true for models with global supersymmetry
(indecomposable representations are common place in
superalgebras). The
Jordan cell structure also appears in the definition of LCFTs. 

Another field where $c=0$ theories are relevant are stochastic
models if the dynamical critical exponent $z=1$. This
possibility was suggested by numerical calculations in a
3-state model at the spinoidal point~\cite{ArndHR98}. In
stochastic models the groundstate is again zero for any number of
sites and, this time, the groundstate wave function has a direct
physical meaning: it describes the probability distribution of the
stationary state. 

In unrelated developments, study~\cite{FridSZ99} of the XXZ chain has lead to
various classes of alternating sign matrices (ASM)~\cite{Bres99} --- a subject of
deep interest in combinatorics. Initially the ASM numbers where found
in the periodic chain ($L$ odd)~\cite{RazuS01a} and have subsequently also been
found for open~\cite{BatcGN01} and twisted boundaries~\cite{BatcGN01,RazuS01b}. They also appear
in the ice model with domain wall boundaries~\cite{Kore82} as was shown in
\cite{Kupe96,RazuS01c}. This model is also related to the combinatorial problem of
domino tiling~\cite{JockPS98}.

A relevant question is --- why should number theoretical properties of wave functions in
quantum mechanics be significant? 
In this letter we answer this question by showing that these ASM numbers are
related to directly measurable quantities. 
Specifically, we bring under one roof some of the
above topics and relate the Hamiltonian (\ref{Ham}) to
stochastic processes on TL algebras where the stationary state probability distribution is
given by the groundstate wavefunction $\ket0$.
A much more detailed version of this work 
will be published elsewhere~\cite{GierPR01}.

\vspace{-.1in}
\section{Temperley-Lieb Algebra}

The generators $\{e_j, j=1,2,\ldots,L\!-\!1\}$ of the TL algebra can be represented
graphically using monoids~\cite{Kauf} 
\bea
&&e_j\;=\punit1\vert 0{-2}1\vert 4{-2}2\punit6\ldots\punit2
  \vert 0{-2}{j\!-\!1}\monoid 4{-2}{j}{j\!+\!1}
  \vert {12}{-2}{j\!+\!2}\punit{14}\ldots\punit2\vert 0{-2}{L-1}\vert 4{-2}{L}\punit8
\eea
The number of independent words $C_L$ in the TL algebra with $L\!-\!1$ generators
is given by the Catalan numbers 
\be
C_n
={1\over n+1}{2n\choose n}=1,2,5,14,\ldots \  n=1,2,3,4\ldots
\ee

We write $e_j=v_{j,j+1}v_{j,j+1}^T$ and 
move to the left ideal by ignoring the upper half of
the monoid diagrams
\be
\begin{array}{c}
\mbox{}\hspace{-25pt}
\monoid0012\punit8\monoid0034\punit5\mapsto\punit1
\downmonoid0012\punit8\downmonoid0034\mbox{}\punit5=v_{1,2}v_{3,4}=\punit1\mbox{}
\put(0,0){\line(1,1){3}}\put(3,3){\line(1,-1){3}}
\put(6,0){\line(1,1){3}}\put(9,3){\line(1,-1){3}}
\put(0,-1){\makebox(0,0)[t]{\smaller \mbox{$0$}}}
\put(3,-1){\makebox(0,0)[t]{\smaller \mbox{$1$}}}
\put(6,-1){\makebox(0,0)[t]{\smaller \mbox{$2$}}}
\put(9,-1){\makebox(0,0)[t]{\smaller \mbox{$3$}}}
\put(12,-1){\makebox(0,0)[t]{\smaller \mbox{$4$}}}
\punit4\\[4pt]
\mbox{}\hspace{-25pt}
\vert0{-2}{1}\punit4
\monoid 0{-2}{2}{3}\monoid {-4}{2}{}{}\punit4
\monoid 0{2}{}{}\punit4\vert 0{-2}{4}
\punit1\mapsto\punit5
\down2monoid 00{1}{4}\punit8\downmonoid {-8}0{2}{3}
\mbox{}\punit1=v_{1,4}v_{2,3}=\punit1\mbox{}
\put(0,0){\line(1,1){3}}
\put(3,3){\line(1,-1){3}}
\put(6,0){\line(1,1){3}}
\put(9,3){\line(1,-1){3}}
\put(3,3){\line(1,1){3}}\put(6,6){\line(1,-1){3}}
\put(0,-1){\makebox(0,0)[t]{\smaller \mbox{$0$}}}
\put(3,-1){\makebox(0,0)[t]{\smaller \mbox{$1$}}}
\put(6,-1){\makebox(0,0)[t]{\smaller \mbox{$2$}}}
\put(9,-1){\makebox(0,0)[t]{\smaller \mbox{$3$}}}
\put(12,-1){\makebox(0,0)[t]{\smaller \mbox{$4$}}}\punit4
\end{array}
\ee
The vector spce generated by this ideal is equivalent to the state space of the $O(1)$
loop model~\cite{BatcGN01}.  Let us assume that $L$ is even. Then the number of
independent words in the left ideal is
$C_{L/2}$. 
The diagrams give the action of the TL generators on the ideal
\bea 
\rule{0pt}{28pt}
e_j(v_{j-1,j+2}v_{j,j+1})\punit2&=&\punit6
\vert{-4}0{j\!-\!1}\monoid00{j}{j\!+\!1}
\down2monoid 04{}{}\punit8\downmonoid {-8}4{}{}\vert00{j\!+\!2}
\mbox{}\punit2=\punit6
\down2monoid 00{j\!-\!1}{j\!+\!2}\punit8\downmonoid {-8}0{j}{j\!+\!1}\nonumber\\
&=&\;v_{j-1,j+2}v_{j,j+1}\punit6\mbox{}\\[20pt]
e_{j-1}(v_{j-1,j+2}v_{j,j+1})\punit2&=&\punit6
\monoid{-4}0{j\!-\!1}{j}\vert{4}0{j\!+\!1}
\down2monoid 04{}{}\punit8\downmonoid {-8}4{}{}\vert00{j\!+\!2}
\mbox{}\punit2=\punit2
\downmonoid00{j\!-\!1}{j}\punit8\downmonoid00{j\!+\!1}{j\!+\!2}\nonumber\\
&=&\;v_{j-1,j}v_{j+1,j+2}
\eea

It is convenient to encode the words in the left ideal by restricted
solid-on-solid (RSOS or Dyck) paths
$\ket a=(a_0,a_1,\ldots,a_{L/2})$ where $a_j$ is the number of half-loops above the midpoint
between sites $j$ and $j\!+\!1$ and $a_{j+1}-a_j=\pm 1$ for each $j$.
For $L=6$ the RSOS paths are

\newpage
\bea
&&L=6:\quad\mbox{\small$\{(0,1,2,3,2,1,0),(0,1,2,1,2,1,0),$}\\
&&\mbox{\small$\ 
(0,1,0,1,2,1,0),(0,1,2,1,0,1,0),(0,1,0,1,0,1,0)\}$}\nonumber
\eea
In this basis for the TL ideal, the $C_{L/2}\times C_{L/2}$ matrix representative of
$H$ for $L=6$ is
\bea
H&\!=\!&-H_{\text{XXZ}}=-\sum_{j=1}^{L-1} (1-e_j)=
\smat{-4\!&\!1\!&\!0\!&\!0\!&\!0\cr 2\!&\!-3\!&\!1\!&\!1\!&\!0\cr 0\!&\!1\!&\!-3\!&\!0\!&\!1\cr
0\!&\!1\!&\!0\!&\!-3\!&\!1\cr 2\!&\!0\!&\!2\!&\!2\!&\!-2}
\label{H6}
\eea
Since $(1\!-\!e_j)=(1\!-\!e_j)^2$ are projectors
we see that $H$ is nonpositive definite. 
The matrix $H$ is representation independent. 
It is straightforward to reconstruct the corresponding eigenvectors of $H_{\text{XXZ}}$ using 
\be
v_{j,k}\;=\;\downmonoid00jk\punit6=q^{-1/2}\ket{\uparrow}_j\!\!\otimes\!\ket{\downarrow}_k-
q^{1/2}\ket{\downarrow}_j\!\!\otimes\!\ket{\uparrow}_k
\ee
where $q=e^{\pi i/3}$ and $\ket{\uparrow}=(1,0)$, $\ket{\downarrow}=(0,1)$ are the spin basis
vectors. Moving to the left ideal changes the degeneracies in the
spectra but not the groundstate wave function. 

\vspace{-.15in}
\section{Temperley-Lieb Stochastic Processes}
It is easy to show that for even $L$ the Hamiltonians $H$ are intensity matrices 
(see \cite{AlcaDR98} for some properties) satisfying $H_{ab}>0$ and $H_{aa}=-\sum_{b\ne a} H_{ab}$. 
The occurence of intensity matrices reveals a novel connection with a stochastic process with
time evolution given by the Euclidean Schr\"odinger equation~\cite{KadaS68}
\be
{d\over dt}P_a(t)=\sum_b H_{ab} P_b(t)
\ee
where $P_a(t)$ is the (unnormalized) probability of finding the system in the state (RSOS path)
$\ket a$ at time $t$. 
In \cite{GierPR01} we will exhibit the rules giving the transition rates $H_{ab}$ for the general
case which define RSOS growth models (for a review of growth models see \cite{HalpZ95}). 
These rates 
$H_{ab}=0,1,2$ for $a\ne b$ are simple. 
At each time step adsorption occurs
with rate $H_{ab}=1$ and desorption with rate $H_{ab}=1,2$. These transition processes
correspond to the addition of a single (diamond shape) tile to the RSOS path or the removal
of a partial layer of tiles and are indicated above and below the main
diagonal respectively in (\ref{H6}). Notice that the desorption process is nonlocal.

\samepage{
Since $H$ is an intensity matrix it has a zero eigenvalue with a trivial bra
\be
\bra0\,H=0,\qquad \bra0=(1,1,\ldots,1)\hspace{1.0in}\mbox{}
\ee
and nontrivial ket wave functions
giving the unnormalized probabilities of the unique stationary state
\be
H\ket0=0,\qquad \ket0=\sum_a P_a\ket a,\quad P_a=\lim_{t\to\infty} P_a(t).
\ee
}

\newpage
\noindent
The wavefunction $\ket0$ describes a critical statistical system of weighted RSOS paths. 
The normalized probabilities are $p_a=P_a/S_1(L)$ where $S_1(L)=\bra0 0\rangle$.
We also have ${}^{T}\hspace{-2pt}\bra00\rangle = S_1(L)^2$
and $\bra00\rangle^T = C_{L/2}$.  
The conjectured~\cite{BatcGN01,RazuS01c} magic properties of the groundstate wave functions
of $H_{\text{XXZ}}$ now assume a direct physical significance.

We have calculated the wavefunctions exactly up to $L=18$. If we now normalize the wave functions 
$\ket0$ to have smallest entry 1
\be
L=6:\quad \ket0=(1,4,5,5,11),\quad S_1(6)=26
\ee
we confirm~\cite{BatcGN01} that the
normalization factors $S_1(L)=\bra0 0\rangle$ satisfy
\bea
S_1(2n)&=&A_V(2n+1)=1,3,26,646,\ldots
\eea
where $A_V(2n\!+\!1)$ is the number of vertically symmetric $\mbox{$(2n\!+\!1)$}\times
(2n\!+\!1)$  ASMs~\cite{Kupe96,Bres99}
\bea
A_V(2n+1)&=& \prod_{j=0}^{n-1}\, (3j+2) {(2j+1)!(6j+3)! \over
(4j+2)!(4j+3)!}
\eea
We also confirm that for the highest path  $P_a=1$ and for the lowest path $P_a=N_8(L)$
where
$N_8$ is the number of cyclically symmetric transpose complement plane
partitions~\cite{Kupe96,Bres99}
\bea
N_8(2n)&=&\prod_{j=0}^{n-1}\,(3j+1)\,{(2j)!(6j)!\over (4j)!(4j+1)!}
\eea

More generally, and to accommodate odd $L$, we can allow $2s$ or fewer defects represented by
unpaired (vertical) lines where
$s=0,\half,1,\ldots,\lfloor L/2\rfloor$ is the spin.  The number of independent words in the
left ideal is then $C_{(L+2s)/2}$. For example, for $L=5$
\bea
\monoid0012\punit8\vert003\punit4\monoid0045\punit5&\mapsto&\punit1
\downmonoid0012\punit8\vertl0032\punit4\downmonoid0045\mbox{}\punit4\;
=\mbox{}
\put(0,0){\line(1,1){3}}\put(3,3){\line(1,-1){3}}
\put(12,0){\line(1,1){3}}\put(15,3){\line(1,-1){3}}
\put(0,-1){\makebox(0,0)[t]{\smaller \mbox{$0$}}}
\put(3,-1){\makebox(0,0)[t]{\smaller \mbox{$1$}}}
\put(6,-1){\makebox(0,0)[t]{\smaller \mbox{$2$}}}
\punit9\vertl0033\punit2
\put(3,-1){\makebox(0,0)[t]{\smaller \mbox{$4$}}}
\put(6,-1){\makebox(0,0)[t]{\smaller \mbox{$5$}}}
\put(9,-1){\makebox(0,0)[t]{\smaller \mbox{$6$}}}
\punit8
\eea
Under the action of the TL generators, the defects can hop and adjacent defects can be
annihilated
\be
\rule{0pt}{12pt}
e_jv_{j-1,j}\;=\punit2\vert00{j\!-\!1}\downmonoid04{}{}\monoid40j{j\!+\!1}\vertl84{}2\punit{10}
=\punit2\vertl00{j\!-\!1}2\downmonoid40j{j\!+\!1}\punit{10}
=v_{j,j+1}\punit2
\ee
The RSOS paths for $L=5, 2s=1$ and $L=4,2s=2$ are
\bea
&&L=5, 2s=1:\quad\mbox{\small$\{(\,|\,0,1,2,1,0),(0,1,2,1,0,1\,|\,),$}\nonumber\\
&&\mbox{\small$\qquad\quad
(\,|\,0,1,0,1,0),(0,1,0,1,0\,|\,),(0,1,0\,|\,0,1,0)\}$}\\
&&L=4,2s=2:\quad\mbox{\small$\{(|\;|\,0,1,0),(|\,0,1,0\,|),(0,1,0\,|\;|),$}\nonumber\\
&&\qquad\qquad\qquad\mbox{\small$\ 
(0,1,2,1,0),(0,1,0,1,0)\}$}
\eea
The matrix representatives of $H$ and wave functions are
\bea
H&\!\!=\!\!\!\!&
\smat{-3\!&\!0\!&\!1\!&\!0\!&\!0\cr 0\!&\!-3\!&\!0\!&\!1\!&\!0\cr
2\!&\!1\!&\!-2\!&\!0\!&\!1\cr 1\!&\!2\!&\!0\!&\!-2\!&\!1\cr 0\!&\!0\!&\!1\!&\!1\!&\!-2},\!
\smat{-2\!&\!1&\!0\!&\!\!\!|\!\!\!\!&\!0\!&\!0\cr
1\!&\!-2\!&\!1\!&\!\!\!|\!\!\!\!&\!0\!&\!0\cr 
\vspace{-2pt} 
0\!&\!1\!&\!-2\!&\!\!\!|\!\!\!\!&\!0\!&\!0\cr 
\vspace{-2pt}
--\!\!\!\!&\!\!\!\!--\!\!\!\!&\!\!\!\!--\!\!\!\!&\!\!\!\!\!\!\!\!&\!\!\!\!--\!\!\!\!&\!\!\!\!--\cr
0\!&\!0\!&\!0\!&\!\!\!|\!\!\!\!&\!-2\!&\!1\cr 
1\!&\!0\!&\!1\!&\!\!\!|\!\!\!\!&\!2\!&\!-1}\mbox{}\hspace{-10pt}\mbox{}
\label{H54}\\
\ket0&=\!&\cases{
(1,1,3,3,3),&$S_1(5)=11,\,  L=5,\, 2s=1$\cr
(0,0,0,1,2),&$S_1(4)=\phantom{2}3,\,  L=4,\, 2s=2$}\mbox{}\hspace{-12pt}\mbox{}
\eea
The matrices are again nonpositive definite intensity matrices. 
For two or more defects the Hamiltonian is (Jordan cell) block triangular and the spectrum decomposes
according  to the $s\ell(2)$ fusion rules which for 2 defects is 
$\half\otimes\half=1\oplus 0$. For $L\le 17$ odd and $2s=1$ we have
verified~\cite{BatcGN01} that
\bea
S_1(2n-1)&=&N_8(2n)=1,2,11,170,7429,\ldots
\eea

It is sometimes convenient to remove the defects by adding $2s$ sites on the right, joining the
defects to these sites by half-loops and working with extended RSOS paths on $L+2s$ sites without
defects. In this scheme the defects are incorporated as boundary conditions. 

As for the combinatorial significance we conjecture~\cite{GierPR01}, following
\cite{RazuS01c}, that for both even and odd $L$ the integers appearing in the groundstate
are given by the number of configurations of the fully packed loop model on a 
$(2L-1)\times(\lfloor L/2\rfloor+1)$ pyramid grid domain with specified 
boundary conditions and links determined by the diagrams in the left ideal. For example, for
$L=6$ and
$v_{1,2}v_{3,4}v_{5,6}$ there are 11 configurations of the type
\be
\punit4
\begin{picture}(20,9)(10,0)
\put(2.7,-.2){.}\put(8.7,-.2){.}
\put(14.7,-.2){.}\put(20.7,-.2){.}\put(26.7,-.2){.}
\put(0,0){\line(0,1){3}}\put(0,3){\line(1,0){3}}\put(3,3){\line(0,1){3}}
\put(3,6){\line(1,0){3}}\put(6,6){\line(0,1){3}}\put(6,9){\line(1,0){3}}
\put(9,6){\line(0,1){3}}\put(9,3){\line(0,1){3}}\put(6,3){\line(1,0){3}}
\put(6,0){\line(0,1){3}}
\put(12,0){\line(0,1){3}}\put(12,3){\line(1,0){6}}\put(18,0){\line(0,1){3}}
\put(12,6){\line(1,0){3}}\put(12,9){\line(1,0){3}}
\put(12,6){\line(0,1){3}}\put(15,6){\line(0,1){3}}
\put(24,0){\line(0,1){3}}\put(21,3){\line(1,0){3}}\put(18,6){\line(1,0){3}}
\put(18,9){\line(1,0){6}}\put(24,6){\line(1,0){3}}\put(27,3){\line(1,0){3}}
\put(18,6){\line(1,0){3}}
\put(21,3){\line(0,1){3}}\put(18,6){\line(0,1){3}}
\put(24,6){\line(0,1){3}}\put(27,3){\line(0,1){3}}
\put(30,0){\line(0,1){3}}
\put(0,-1){\makebox(0,0)[t]{\smaller \mbox{$1$}}}
\put(6,-1){\makebox(0,0)[t]{\smaller \mbox{$2$}}}
\put(12,-1){\makebox(0,0)[t]{\smaller \mbox{$3$}}}
\put(18,-1){\makebox(0,0)[t]{\smaller \mbox{$4$}}}
\put(24,-1){\makebox(0,0)[t]{\smaller \mbox{$5$}}}
\put(30,-1){\makebox(0,0)[t]{\smaller \mbox{$6$}}}
\end{picture}
\punit2\sim\punit2
\downmonoid0012\punit8\downmonoid0034\punit8\downmonoid0056
\ee

\vspace{-.15in}
\section{Properties of Stationary States}

The stationary states are independent of the initial condition and Jordan cell stucture but the time
evolution does depend on  which Jordan blocks are selected out by the initial $t=0$ probabilities.
A stochastic process can also be defined by the action of $H$ on the
whole TL algebra ($C_L$ states) giving a different Jordan cell structure and (non-unique) stationary
states.

Many properties of the stationary states can be studied by finite-size scaling.
We give preliminary results on the average perimeter $\langle\ell(a)\rangle$, average area
$\langle N(a)\rangle$ and average number of clusters $\langle C(a)\rangle$ where we define
\bea
&&\ell(a)=\!\!\!\!\sum_{j=1}^{2\lceil L/2\rceil\!-\!1}\!\!\! \half|a_{j\!+\!1}-a_{j\!-\!1}|,\quad
N(a)=\!\!\!\sum_{j=1}^{2\lceil L/2\rceil} \lfloor a_j/2\rfloor\\
&&C(a)=\sum_{j=1}^L \delta(a_j,0)-2s,\qquad \langle\ldots\rangle=\sum_a \ldots p_a
\eea
Here $N(a)$ is the number of tiles and we have ignored the fixed horizontal contribution 
$2\lceil L/2\rceil$ to the perimeter. 

\newpage
\mbox{}\vspace{-.3in}

By extrapolating results for $L\le 18$ we find numerically the following 
scaling behaviour for large even $L$
\bea
\langle\ell(a)\rangle&\!\sim\!&.249(1)\,L,\,
\langle N(a)\rangle\!\sim\!.065(1)\,L^{1+\nu}\!,\,\nu=.50(3)\\
&&\qquad \langle C(a)\rangle\sim 1.17 (1)\;L^x,\ x=.667(3)
\eea
The same exponents but different amplitudes are obtained for $\langle\ell(a)\rangle$ and $\langle
N(a)\rangle$ with odd $L$. The finite-size behaviour of other properties describing the critical
weighted RSOS paths model and its connection to conformal invariance will be discussed in
\cite{GierPR01}. 

\vspace{-.05in}
\section{Conformally Invariant Spectra}

We assert that the spectra of the intensity matrices $H$ are described by a conformal field
theory~\cite{ReadS01} with central charge $c=0$, conformal weights 
\be
\Delta_s=\mbox{${s(2s-1)\over 3}=0,0,{1\over 3},1\ldots\quad s=0,{1\over 2},1,{3\over 2},\ldots$}
\ee
and Virasoro characters with $q$ the modular parameter
\be
\chi_s(q)=q^{\Delta_s}\big(1-q^{2s+1}\big)\prod_{n=1}^\infty (1-q^n)^{-1}
\ee

Finite-size corrections~\cite{Card86} to the energy levels $E_n$, $H\ket n=-E_n\ket n$, for large
$L$ with $n=0,1,2,\ldots$ are 
\be
{LE_n\over \pi v}=\Delta_s+k_n+o(1),\quad k_n\in {\Bbb N}
\ee
where $k_n$ labels descendents and $v=3\sqrt{3}/2$~\cite{ABBBQ} is the sound velocity.
We have calculated numerically the finite-size spectra up to size $L=16$ and obtain the following
estimates using Van den Broeck-Schwartz~\cite{VBS} approximants.

\begin{table}[h]
\begin{center}
\begin{tabular}{l|ll|ll|ll}
&\multicolumn{2}{c}{$s=0$}&\multicolumn{2}{c}{$s=\half$}&\multicolumn{2}{c}{$s=1$}\\
$n$&$\Delta_s\!+\!k_n$&exact&$\Delta_s\!+\!k_n$&exact&$\Delta_s\!+\!k_n$&exact\\
\hline
0&0&0&0&0&0.333(1)&1/3\\
1&1.999(1)&2&1.0&1&1.34(2)&4/3\\
2&3.003(4)&3&1.999(3)&2&2.4(2)&7/3\\
3&4.01(2)&4&3.003(6)&3&2.3(2)&7/3\\
4&3.99(8)&4&2.999(6)&3&3.4(3)&10/3\\
5&4.7(5)&5&3.8(5)&4&3.5(4)&10/3
\end{tabular}
\end{center}
\caption{Table of energy level estimates for
$2s=0,1,2$ defects giving the characters $\chi_s(q)$ for
$s=0,\half,1$.}
\end{table}
\mbox{}\vspace{-.2in}\mbox{}
\section*{Acknowledgements} This research is supported by the Australian Research Council. VR is
an ARC IREX Fellow. We thank Murray Batchelor, Mikhail Flohr, Bernard Nienhuis and Aleks
Owczarek for discussions. 

\newpage
\samepage{
\mbox{}\vspace{-.55in}\mbox{}

}

\end{document}